\documentclass[journal,final,letterpaper,twoside,twocolumn]{IEEEtran}

\usepackage{dsfont}
\usepackage{amsmath}
\usepackage{amssymb}
\usepackage{amsfonts}
\usepackage{graphicx}
\usepackage{amsmath}
\usepackage[all,poly]{xy}
\usepackage{multirow}
\usepackage{algorithm}
\usepackage{algorithmic}
\usepackage{color}
\usepackage[noadjust]{cite}
\usepackage{array}
\usepackage{url}

\newcommand{\figwidth}{\columnwidth}

\newcommand{\nosou}{r}\newcommand{\nbsou}{R}\newcommand{\noband}{l}\newcommand{\nbband}{L}\newcommand{\noobs}{p}\newcommand{\nbobs}{P}\newcommand{\MATobs}{\mathbf X}\newcommand{\Vobs}[1]{{\mathbf x}_{#1}}\newcommand{\MATsou}{\mathbf S}\newcommand{\Vsou}[1]{{\mathbf s}_{#1}}\newcommand{\sou}[2]{s_{#1,#2}}\newcommand{\abond}[2]{a_{#1,#2}}\newcommand{\Vabond}[1]{{\bold a}_{#1}}\newcommand{\MATabond}{\mathbf A}
\newcommand{\Vnoisevar}{{\boldsymbol \sigma}_{\mathrm{e}}^2}\newcommand{\noisevar}[1]{\sigma_{\mathrm{e},#1}^2}
\newcommand{\soucoeffA}[1]{\alpha_{#1}}\newcommand{\soucoeffB}[1]{\beta_{#1}}\newcommand{\abondcoeffA}[1]{\lambda_{#1}}\newcommand{\abondcoeffB}[1]{\gamma_{#1}}

\newcommand{\transp}{^{\mathsf T}}\newcommand{\Simplex}{\mathbb{S}}\newcommand{\Indicfun}[2]{\textbf{1}_{#1}(#2)}

\begin{document}

\title{Implementation strategies for hyperspectral unmixing using Bayesian
source separation}

\author{Fr\'ed\'eric Schmidt, Albrecht Schmidt, Erwan Tr\'eguier,
\\ Ma\"el Guiheneuf, Sa\"id Moussaoui and Nicolas
Dobigeon,~\IEEEmembership{Member,~IEEE,}
\thanks{Fr\'ed\'eric Schmidt was with ESA, ESAC, Villanueva de la Canada, Madrid, Spain. He is now with University Paris-Sud, Laboratoire IDES, UMR8148, Orsay, F-91405, CNRS, Orsay, F-91405 (e-mail:
frederic.schmidt@u-psud.fr).}
\thanks{Albrecht Schmidt, Erwan Tr\'eguier and Ma\"el Guiheneuf are with ESA, ESAC, Villanueva de la Canada, Madrid, Spain (e-mail:
Albrecht.Schmidt@esa.int, mael.guiheneuf@gmail.com,
erwan.treguier@sciops.esa.int).}
\thanks{Sa\"id Moussaoui is with IRCCyN, UMR CNRS 6597, \'Ecole Centrale Nantes, France (e-mail:
said.moussaoui@irccyn.ec-nantes.fr).}
\thanks{Nicolas Dobigeon is with the University of Toulouse,
IRIT/INP-ENSEEIHT/T\'eSA, 2 rue Charles Camichel, BP 7122, 31071
Toulouse cedex 7, France (e-mail: Nicolas.Dobigeon@enseeiht.fr).}}

\maketitle

\begin{abstract}
Bayesian Positive Source Separation (BPSS) is a useful unsupervised
approach for hyperspectral data unmixing, where numerical
non-negativity of spectra and abundances has to be ensured, such in
remote sensing. Moreover, it is sensible to impose a sum-to-one
(full additivity) constraint to the estimated source abundances in
each pixel. Even though non-negativity and full additivity are two
necessary properties to get physically interpretable results, the
use of BPSS algorithms has been so far limited by high computation
time and large memory requirements due to the Markov chain Monte
Carlo calculations. An implementation strategy which allows one to
apply these algorithms on a full hyperspectral image, as typical in
Earth and Planetary Science, is introduced. Effects of pixel
selection, the impact of such sampling on the relevance of the
estimated component spectra and abundance maps, as well as on the
computation times, are discussed. For that purpose, two different
dataset have been used: a synthetic one and a real hyperspectral
image from Mars.
\end{abstract}

\begin{keywords}
Hyperspectral imaging, source separation, Bayesian estimation, implementation
strategy, computation time.
\end{keywords}

\section{Introduction}

\label{sec:introduction}

In visible and near infrared hyperspectral imaging, each image
recorded by the sensor is the solar light reflected and diffused
back from the observed planet surface and atmosphere at a particular
spectral band. Under some assumptions related to surface and
atmosphere properties -- e.g., Lambertian surface, no intimate
mixture, no diffusion terms in the atmosphere, homogeneous geometry
in the scene -- each measured spectrum -- i.e., each pixel of the
observed image for several spectral bands -- is modeled as a linear
mixture of the scene component spectra (\emph{endmembers})
\cite{Tanre_modeling_groundreflectance_AppliedOptics1979,Healy99,Keshava2002}.
In this model, the weight of each component spectrum is linked to
its abundance in the surface area which corresponds to the
underlying pixel. The main goal of hyperspectral unmixing is to
identify the components of the imaged surface and to estimate their
respective abundances \cite{Schott97,Chang07}.

By considering $P$ pixels of an hyperspectral image acquired in $L$
frequency bands, the observed spectra are gathered in a $P\times L$
data matrix $\mathbf{X}$, potentially ignoring spatiality. Each row
of this matrix contains a measured spectrum at a pixel with spatial
index $p=1,\ldots,P$. According to the linear mixing model, the
$p$th spectrum, $1\leq p\leq P$, can be expressed as a linear
combination of $R$ pure spectra of the surface components. Using
matrix notations, this linear spectral mixing model can be written
as
\begin{equation}
\mathbf{X}\approx\mathbf{A}\mathbf{S}
\end{equation}
where non-negative matrices $\mathbf{A}\in \mathbb{R}_+^{P\times R}$
and $\mathbf{S}\in \mathbb{R}_+^{R\times L}$ approximate
$\mathbf{X}\in \mathbb{R}_+^{P\times L}$ in the sense that
$\frac{1}{2}||\mathbf{A}\mathbf{S}-\mathbf{X}||^{2}$ is minimized
($\mathbb{R}_+^{\cdot\times\cdot}$ denotes the space of matrices
with only non-negative entries of respective dimensions). The rows
of matrix $\mathbf{S}$ now contain the pure surface spectra of the
$R$ components, and each element $a_{pr}$ of matrix $\mathbf{A}$
corresponds to the abundance of the $r$th component in pixel with
spatial index $p$. For a qualitative and quantitative description of
the observed scene composition, the estimation problem consists of
finding matrices $\mathbf{S}$ and $\mathbf{A}$ that allow one to
explain the data matrix $\mathbf{X}$ and have a coherent physical
interpretation. This approach casts the hyperspectral unmixing as a
\emph{source separation} problem under a linear instantaneous mixing
model \cite{Comon91}. Source separation is a statistical
multivariate data processing problem whose aim is to recover unknown
signals (called \emph{sources}) from noisy and mixed observations of
these sources \cite{Cichocki02,HyvarinenBOOK2001}.

This problem has been studied in-depth in recent years, starting
with pioneer work more than 15 years ago \cite{PT94,LS99}. From a
statistical point of view, the problem is also related to Principal
Component Analysis (PCA) and k-means clustering (see \cite{B+07} for
an overview). Also note that the factorization
$\mathbf{A}\mathbf{S}$ is not uniquely defined. For instance, for
any matrices $\mathbf{Z}\in \mathbb{R}_+^{R\times.}$ such that
$\mathbf{Z}  \mathbf{Z}^{-1}=\mathbf{I}$, then
$\mathbf{A}\mathbf{Z}\mathbf{Z}^{-1}\mathbf{S}=(\mathbf{A}\mathbf{Z})(\mathbf{Z}^{-1}\mathbf{S})=\mathbf{A}'\mathbf{S}'$
is a solution as well; this holds even if the minimization is able
to find a global minimum. However, when solving this separation
problem with hyperspectral data, several constraints can be
considered to reduce the set of admissible solutions. A first hard
constraint is the non-negativity of the elements of both matrices
$\mathbf{S}$ and $\mathbf{A}$ since they correspond to pure spectra
and abundances of the surface components, respectively. A second
constraint that may be imposed is the sum-to-one (additivity)
constraint of the abundances. Indeed the abundance weights
correspond to proportions and therefore should sum to unity.

Several algorithms have been proposed in the literature to solve
fully constrained unmixing problems, i.e., handling both of the
constraints imposed on the spectra and abundances. Specifically, an
iterative algorithm called ICE (Iterated Constrained Endmembers) has
been proposed in \cite{Berman2004} to minimize the size of the
simplex formed by the estimated endmembers. However, as noted in
\cite{Nascimento2007igarss}, results provided by ICE strongly depend
on the choice of the algorithm parameters. More recently, Jia and
Qian have developed in \cite{Jia2007} complexity-based BSS
algorithms that exploit the pixel correlations to recover endmember
signatures. In \cite{Miao2007}, Miao et al. have introduced a
non-negative matrix factorization (NMF) algorithm with an additivity
penalty on the abundance coefficients. Similarly, other constrained
NMF approaches exploiting smoothness and sparseness features have
been considered in \cite{JiaQian4694061}. Note that all the
strategies described above are based on an optimization scheme to
minimize a penalty criterion. Consequently, they may suffer from
convergence issues, e.g., due to the presence of local maxima and
the large number of parameters to be estimated.

Alternatively, the constrained separation problem can be
conveniently addressed in a Bayesian framework. Two algorithms that
perform unsupervised separation under positivity and sum-to-one
constraints have been recently proposed
\cite{Moussaoui06tsp,Dobigeon09sp}. These algorithms are based on
hierarchical Bayesian modeling to encode prior information regarding
the observation process, the parameters of interest and include the
positivity and full additivity constraints. The complexity of the
inference from the posterior distribution of the parameters of
interest is tackled using Markov chain Monte Carlo (MCMC) methods
\cite{Gilks96,Gelfand90}, which has been proposed to analyze
hyperspectral images \cite{Bali_Hyperspectral_ImagProces2008}. The
algorithmic details are not described here. The reader is invited to
read the pseudo-codes summarized in Algo.'s 1 and 2, and to consult
\cite{Moussaoui06tsp} and \cite{Dobigeon09sp} for further details.
The only difference between the two BPSS algorithms is the sampling
of the abundance vectors $\Vabond{\noobs}$
($\noobs=1,\ldots,\nbobs$). However, since these algorithms rely on
MCMC methods, the computation time drastically increases with the
image size and these algorithms have not been applied for large
scale data processing in spite of their high effectiveness.

The aim of this article is to discuss some implementation strategies
which allow one to apply these algorithms to real hyperspectral
data, even if images are large. Previous works about blind source
separation of hyperspectral images have been proposed
\cite{Naceur_SourceSeparation_IEEE_2004,Nascimento_ICAandTransfertRadia_IEEE_2005,Wang_ICAEndmemberAbundnace_TGRS2006}
but only few use positivity/sum-to-unity constraints
\cite{Moussaoui08neuro}. To overcome this difficulty, a first
approach has been proposed in \cite{Moussaoui08neuro} to combine
Independent Component Analysis (ICA) and Bayesian Positive Source
Separation (BPSS). Firstly, applying an ICA algorithm (such as JADE
\cite{Cardoso93} or FastICA\cite{Hyvarinan97}) to the hyperspectral
images is applied to get a rough spatial classification of the scene
and to sample relevant pixels (i.e., from each \emph{class}, the
pixels whose spectra are mostly uncorrelated are selected).
Secondly, the spectra associated to these pixels will serve in the
Bayesian separation algorithm to estimate the endmember spectra.
Finally, the abundances can then be estimated on the whole image
using the estimated spectra. However, this strategy presents a
limitation related to the difficulty to determine the number of
pixels to retain from each independent component class. In this
paper, another pixel selection strategy based on the computation of
the convex hull of the hyperspectral data is introduced. Its
influence on the separation performances is also discussed. The
issue of estimating the number of sources, or ``intrinsic
dimension''
\cite{Chang_NumberOFpolesEstimation_IEEEGeoRemSens_2004}, will not
be addressed in this article. Several methods have been proposed in
the literature
\cite{Bioucas-Dias_HyperspecSubspaceIdentif_IEEEtgrs2008,Luo_UnsupervisedClassif_Conf2009}.

This paper is organized as follows. Section
\ref{sec:optimization_strategies} describes the proposed
implementation strategies adopted for this work. Section
\ref{sec:TechnicalOptimization} summarizes the improvements related
to the technical aspects of memory storage and computation issues.
Section \ref{sec:PixelSelection} discusses the performances of the
resulting algorithms when the pixel selection preprocessing step is
introduced.

\begin{algorithm}[htpb]
    \scriptsize{
   \label{algo:Gibbs}
   \caption{Bayesian positive source separation algorithm (BPSS)}
    \begin{algorithmic}
     \FOR{$i=1,\ldots,N_{\textrm{MC}}$}
        \STATE \emph{\scriptsize{\% sampling the abundance hyperparameters}}
        \FOR{$\noobs=1,\ldots, \nbobs$}
            \STATE Draw $\abondcoeffA{\noobs}$ from the pdf
                \begin{equation*}
                \label{eq:gene_abondcoeffA}
                f\left(\abondcoeffA{\noobs} \big| \Vabond{\noobs:}, \abondcoeffB{\noobs}\right) \propto
                    \prod_{\nosou=1}^{\nbsou} \left[\frac{\abondcoeffB{\noobs}^{\abondcoeffA{\noobs}}}{\Gamma\left(\abondcoeffA{\noobs}\right)}
                        \abond{\noobs}{\nosou}^{\abondcoeffA{\noobs}}\right]
                        e^{-\epsilon\abondcoeffA{\noobs}}\Indicfun{\mathbb{R}^+}{\abondcoeffA{\noobs}}.
                \end{equation*}
         \ENDFOR
        \STATE \emph{\scriptsize{\% sampling the abundance hyperparameters}}
        \FOR{$\noobs=1,\ldots, \nbobs$}
            \STATE Draw
            $ \abondcoeffB{\noobs}$ from the gamma
            distribution
                \begin{equation*}
                \label{eq:gene_abondcoeffB}
                \abondcoeffB{\noobs} \big|  \abondcoeffA{\noobs}, \Vabond{\noobs:} \sim
                \mathcal{G}\left(1 + \nbsou  \abondcoeffA{\noobs} + \epsilon,\sum_{\nosou=1}^{\nbsou} \abond{\noobs}{\nosou} +
                    \epsilon\right).
                \end{equation*}
        \ENDFOR
        \STATE \emph{\scriptsize{\% sampling the abundance vectors}}
        \FOR{$\noobs=1,\ldots, \nbobs$ and $\nosou=1,\ldots,\nbsou$}
            \STATE Draw
            $\abond{\noobs}{\nosou}$ from the pdf
                \begin{multline*}\label{eq:gene_abond}
                f\left(\abond{\noobs}{\nosou} \big|\abondcoeffA{\noobs},\abondcoeffB{\noobs},\MATsou,\Vnoisevar,\MATobs \right)
                    \\ \propto \abond{\noobs}{\nosou}^{\abondcoeffA{\nosou}-1}
                    \Indicfun{\mathbb{R}^+}{\abond{\noobs}{\nosou}}
                    \exp\left[-\frac{\left(\abond{\noobs}{\nosou}-\mu_{\noobs,\nosou}\right)^2}{2\delta^2_{\noobs}} - \abondcoeffB{\noobs} \abond{\noobs}{\nosou}\right],
                \end{multline*}
        \ENDFOR
        \STATE \emph{\scriptsize{\% sampling the noise hyperparameter}}
        \STATE Draw $\psi_\mathrm{e}$ from the
        inverse-gamma distribution
                \begin{equation*}
                    \label{eq:gene_hypernoise}
                    \psi_\mathrm{e}\left|\Vnoisevar,\rho_\mathrm{e}\right. \sim
                    \mathcal{IG}\left(\frac{\nbobs\rho_\mathrm{e}}{2},\frac{1}{2}\sum_{\noobs=1}^{\nbobs}\frac{1}{\noisevar{\noobs}}\right).
                    \end{equation*}
        \STATE \emph{\scriptsize{\% sampling the noise variances}}
        \FOR{$\noobs=1,\ldots,\nbobs$}
            \STATE Draw $ \noisevar{\noobs}$ from the inverse-gamma distribution
                \begin{equation*} \label{eq:gene_noisevar}
                \noisevar{\noobs}\left|\psi_{\mathrm{e}},\Vabond{\noobs:},\MATsou,\Vobs{\noobs:}\right.
                \sim \mathcal{IG}\left(\frac{\rho_{\mathrm{e}}+\nbband}{2},
                \frac{\psi_{\mathrm{e}}+\left\|\Vobs{\noobs:}-\MATsou\Vabond{\noobs:}\right\|^2}{2}\right).
                \end{equation*}
         \ENDFOR
        \STATE \emph{\scriptsize{\% sampling the source hyperparameters}}
        \FOR{$\nosou=1,\ldots, \nbsou$}
            \STATE Draw $\soucoeffA{\nosou}$ from the pdf
                \begin{equation*}
                \label{eq:gene_soucoeffA}
                f\left(\soucoeffA{\nosou} \big|\Vsou{\nosou:}, \soucoeffB{\nosou}\right) \propto
                    \prod_{\noband=1}^{\nbband} \left[\frac{\soucoeffB{\nosou}^{\soucoeffA{\nosou}}}{\Gamma\left(\soucoeffA{\nosou}\right)}
                        \sou{\nosou}{\noband}^{\soucoeffA{\nosou}}\right]
                        e^{-\epsilon\soucoeffA{\nosou}}\Indicfun{\mathbb{R}^+}{\soucoeffA{\nosou}}.
                \end{equation*}
         \ENDFOR
        \STATE \emph{\scriptsize{\% sampling the source hyperparameters}}
        \FOR{$\nosou=1,\ldots, \nbsou$}
            \STATE Draw
            $ \soucoeffB{\nosou}$ from the gamma
            distribution
                \begin{equation*}
                \label{eq:gene_soucoeffB}
                \soucoeffB{\nosou} \big|  \soucoeffA{\nosou}, \Vsou{\nosou:} \sim
                \mathcal{G}\left(1 + \nbband  \soucoeffA{\nosou} + \epsilon,\sum_{\noband=1}^{\nbband} \sou{\nosou}{\noband} +
                    \epsilon\right).
                \end{equation*}
        \ENDFOR
        \STATE \emph{\scriptsize{\% sampling the source spectrum}}
        \FOR{$\nosou=1,\ldots, \nbsou$ and $\noband=1,\ldots,\nbband$}
            \STATE Draw
            $\sou{\nosou}{\noband}$ from the pdf
                \begin{multline*}
                f\left(\sou{\nosou}{\noband} \big|\soucoeffA{\nosou},\soucoeffB{\nosou},\MATabond,\Vnoisevar,\MATobs \right)\\
                 \propto \sou{\nosou}{\noband}^{\soucoeffA{\nosou}-1}
                    \Indicfun{\mathbb{R}^+}{\sou{\nosou}{\noband}}
                    \exp\left[-\frac{\left(\sou{\nosou}{\noband}-\mu_{\nosou,\noband}\right)^2}{2\delta^2_{\nosou}} - \soucoeffB{\nosou} \sou{\nosou}{\noband}\right],
                \end{multline*}
        \ENDFOR
    \ENDFOR
    \end{algorithmic}}
\end{algorithm}

\begin{algorithm}[htpb]
 \scriptsize{
   \label{algo:Gibbs2}
   \caption{Fully constrained Bayesian positive source separation algorithm (BPSS2)}
    \begin{algorithmic}
     \FOR{$i=1,\ldots,N_{\textrm{MC}}$}
        \STATE \emph{\scriptsize{\% sampling the abundance vectors}}
        \FOR{$\noobs=1,\ldots,\nbobs$}
            \STATE Draw $\Vabond{\noobs:}$ from the pdf
                \begin{multline*}
                    f\left(\Vabond{\noobs:}|\MATabond,\Vnoisevar,\MATobs\right) \\
                   \propto \exp\left[-\frac{1}{2}\left(\Vabond{\noobs:}-\boldsymbol{\mu}_{\noobs}\right)\transp
                   \boldsymbol{\Lambda}_{\noobs}^{-1}\left(\Vabond{\noobs:}-\boldsymbol{\mu}_{\noobs}\right)\right]
                    \Indicfun{\Simplex}{\Vabond{\noobs:}}.
                \end{multline*}
            with
                \begin{equation*}
                \label{eq:space_S2}
                \Simplex=\left\{ \Vabond{\noobs:} ;  \abond{\noobs}{\nosou} \geq 0, \ \forall \nosou =1,\ldots,\nbsou,
                \; \sum_{\nosou=1}^{\nbsou} \abond{\noobs}{\nosou}= 1 \right\}.
                \end{equation*}
         \ENDFOR
        \STATE \emph{\scriptsize{\% sampling the noise hyperparameter}}
        \STATE Draw $\psi_\mathrm{e}$ from the
        inverse-gamma distribution
                \begin{equation*}
                    \label{eq:gene_hypernoise2}
                    \psi_\mathrm{e}\left|\Vnoisevar,\rho_\mathrm{e}\right. \sim
                    \mathcal{IG}\left(\frac{\nbobs\rho_\mathrm{e}}{2},\frac{1}{2}\sum_{\noobs=1}^{\nbobs}\frac{1}{\noisevar{\noobs}}\right).
                    \end{equation*}
        \STATE \emph{\scriptsize{\% sampling the noise variances}}
        \FOR{$\noobs=1,\ldots,\nbobs$}
            \STATE Draw $ \noisevar{\noobs}$ from the inverse-gamma distribution
                \begin{equation*} \label{eq:gene_noisevar2}
                \noisevar{\noobs}\left|\psi_{\mathrm{e}},\Vabond{\noobs:},\MATsou,\Vobs{\noobs:}\right.
                \sim \mathcal{IG}\left(\frac{\rho_{\mathrm{e}}+\nbband}{2},
                \frac{\psi_{\mathrm{e}}+\left\|\Vobs{\noobs:}-\MATsou\Vabond{\noobs:}\right\|^2}{2}\right).
                \end{equation*}
         \ENDFOR
        \STATE \emph{\scriptsize{\% sampling the source hyperparameters}}
        \FOR{$\nosou=1,\ldots, \nbsou$}
            \STATE Draw $\soucoeffA{\nosou}$ from the pdf
                \begin{equation*}
                \label{eq:gene_soucoeffA2}
                f\left(\soucoeffA{\nosou} \big|\Vsou{\nosou:}, \soucoeffB{\nosou}\right) \propto
                    \prod_{\noband=1}^{\nbband} \left[\frac{\soucoeffB{\nosou}^{\soucoeffA{\nosou}}}{\Gamma\left(\soucoeffA{\nosou}\right)}
                        \sou{\nosou}{\noband}^{\soucoeffA{\nosou}}\right]
                        e^{-\epsilon\soucoeffA{\nosou}}\Indicfun{\mathbb{R}^+}{\soucoeffA{\nosou}}.
                \end{equation*}
         \ENDFOR
        \STATE \emph{\scriptsize{\% sampling the source hyperparameters}}
        \FOR{$\nosou=1,\ldots, \nbsou$}
            \STATE Draw
            $ \soucoeffB{\nosou}$ from the gamma
            distribution
                \begin{equation*}
                \label{eq:gene_soucoeffB2}
                \soucoeffB{\nosou} \big|  \soucoeffA{\nosou}, \Vsou{\nosou:} \sim
                \mathcal{G}\left(1 + \nbband  \soucoeffA{\nosou} + \epsilon,\sum_{\noband=1}^{\nbband} \sou{\nosou}{\noband} +
                    \epsilon\right).
                \end{equation*}
        \ENDFOR
        \STATE \emph{\scriptsize{\% sampling the source spectrum}}
        \FOR{$\nosou=1,\ldots, \nbsou$ and $\noband=1,\ldots,\nbband$}
            \STATE Draw
            $\sou{\nosou}{\noband}$ from the pdf
                \begin{multline*}\label{eq:gene_mat2}
                f\left(\sou{\nosou}{\noband} \big|\soucoeffA{\nosou},\soucoeffB{\nosou},\MATabond,\Vnoisevar,\MATobs \right)
                   \\ \propto \sou{\nosou}{\noband}^{\soucoeffA{\nosou}-1}
                    \Indicfun{\mathbb{R}^+}{\sou{\nosou}{\noband}}
                    \exp\left[-\frac{\left(\sou{\nosou}{\noband}-\mu_{\nosou,\noband}\right)^2}{2\delta^2_{\nosou}} - \soucoeffB{\nosou} \sou{\nosou}{\noband}\right],
                \end{multline*}
        \ENDFOR
    \ENDFOR
    \end{algorithmic}}
\end{algorithm}

\section{Optimization Strategies\label{sec:optimization_strategies}}

The optimization consists of two independent parts which will be
referred to: (i) \emph{Technical Optimization (TO)} to reduce the
memory footprint, the average cost of algorithmic operations, and
make smarter reuse of memory (ii) \emph{Convex Hull Optimization
(CHO)} to reduce the number of spectra to be processed.

Both parts enabled us to analyze hyperspectral images that so far
were not open to analysis. The authors stress that the techniques
applied in (i) do not alter the results of the original algorithm
(see section \ref{sec:TechnicalOptimization}). On the other hand,
the optimization strategy (ii) only selects a subset of the original
input and therefore may change the results. Impact of the strategy
(ii) needs to be evaluated, which will be presented in section
\ref{sec:PixelSelection}.

\subsection{Technical Optimization (TO)}

The algorithms introduced in \cite{Moussaoui06tsp,Dobigeon09sp} and
referred to as BPSS and BPSS2, respectively, could be successfully
launched on an image of a restricted size, typically of a few
thousand pixels. The main goal of this work is to optimize the
memory requirement of these algorithms to process a whole
hyperspectral image of $100 000$ spectra as it typically occurs in
Earth and Planetary Science. Since the time requirements of the
computation increase drastically for a larger number of pixels and a
larger number of sources, another challenging objective is to reduce
as much as possible the computation time. In that respect, our
proposal is to discuss the memory storage, the data representation,
the operating system architecture and the computing parallelization.
These algorithms have been implemented in MATLAB$^\copyright$ for
this work but future implementations will be done in other languages
as well.

\subsubsection{Memory}

Thanks to the MATLAB$^\copyright$ profiler, it can be noticed that
the main limitation of the BPSS implementation is the contiguous
memory. Fragmentation may occur when variables are resized after
memory allocation. In this case, the memory management might not be
able to allocate a chunk of memory that is large enough to hold the
new variable. Significant garbage collection may set in, which may
have a significant performance impact. In our case, to reduce the
impact of garbage collection, pre-allocating the matrices and work
with global variables has been found to be useful.

\subsubsection{Precision}

MATLAB$^\copyright$ by default computes on double precision. However
computing with single data type saves a lot of computation time
while providing sufficient arithmetic precision. It has been
estimated to win up to $60\%$ computation time on an x86 processor
architecture, while the changes to the code have been minimal.
Furthermore, most dataset come as single precision.

\subsubsection{OS Architecture}

It is interesting to note that MATLAB$^\copyright$ is limited in
terms of memory usage (regardless of the size of physical memory).
This depends on the Operating System (OS) and on the MATLAB version
(see Table \ref{Flo:TableMemOS}).

\begin{table}
\center\begin{tabular}{|c|c|}
\hline Operating System  & Memory Limitation  \\
\hline
\hline
32-bit Microsoft Windows XP & \multirow{2}{*}{2GB}\\
Windows Vista  & \\
\hline
32-bit Windows XP with 3 GB \emph{boot.ini} switch & \multirow{2}{*}{3GB}\\
32-bit Windows Vista with \emph{increaseuserva} set  &  \\
\hline
32-bit LINUX & 3GB \\
\hline
64-bit Windows XP, Linux, & \multirow{3}{*}{4GB}\\
Apple Macintosh OS X      &\\
or SunSolaris running 32-bit MATLAB$^\copyright$ &   \\
\hline
64-bit Windows XP, Windows Vista, & \multirow{2}{*}{8GB} \\
Linux, or Solaris running 64-bit MATLAB$^\copyright$ &  \\ \hline
\end{tabular}
\label{Flo:TableMemOS} \caption{Summary of memory limitation
depending on Operating System.}
\end{table}

Therefore, a 32-bits LINUX architecture has been chosen.

\subsubsection{Parallelization}

MATLAB$^\copyright$ contains libraries dedicated to automatically
parallelize parts of the algorithms on a single computer. BPSS has
been run on a $4$-core machine. The underlying matrix libraries
already provide a certain level of parallelism depending on the
number of available cores. However, in the future, parts of the code
could be parallelized and the jobs could be submitted to a grid in
order to speed up the calculation process.

\subsection{Convex Hull Optimization (CHO)}

The proposed pixel selection strategy is based on the convex hull of
the data matrix projection into the subspace spanned by the
principal components. The convex hull of a point set is the smallest
convex set that includes all the points \cite{Barber_convexHull}.
The pixels associated to the vertices of the convex hull are
selected \cite{Boardman_1993} and are expected, despite their
limited number, to exhibit the main spectral features of the whole
dataset. In terms of abundances, this sample of points should
contain the pixels with the highest abundances of the components
which contribute to the investigated hyperspectral image (i.e., the
purest pixels or most extreme pixels). It can be used as a concise
representation of the dataset which still features the strongest
spectral signatures available in the original image. This strategy
is also used as a first step in endmember extraction algorithms for
dimension reduction and purest pixel determination
\cite{Chang06,Nascimento05,ifarraguerri99,Plaza04,Craig94,Boardman_1993}.
Pixel selection has the advantage, to reduce the number of mixture
spectra to unmix and to enforce the sparsity of the mixing
coefficients to be estimated. Note that the spectral dimension of
the selected spectra is not changed, only the spatial dimension is
reduced since only few pixels are selected.

The convex hull selection has been implemented after seven spectral
components have been selected through PCA, which turned out to be a
good compromise between resource consumption and accuracy.

\section{Performance and Accuracy of TO}
\label{sec:TechnicalOptimization} All the following runs are
performed on a Quad-Core AMD Opteron(tm) Processor 8384 at $2.7$GHz
with $2$Gb of memory.

\subsubsection{Performance}

Computation times between the previous version of BPSS and the TO
version have been compared when processing a synthetic dataset of
$1052$ spectra of $128$ bands and $3$ sources. For a run attempting
at estimating $3$ sources, the computation time has decreased from
$1106$s (previous version) to $724$s (TO version), i.e., by a factor
of about $1.5$. In addition, the total memory consumption is nearly
half for the TO version of the algorithm.

\subsubsection{Accuracy}

Due to the stochastic nature of the BPSS algorithms, it is difficult
to demonstrate that two algorithms are semantically identical. In
order to check that no significant loss of accuracy has been induced
by the TO and especially by the change from double to single
precision, several tests have been performed with different random
seeds $\chi_{1}$ and $\chi_{2}$, used for the initialization step of
the MCMC. The sources $S_{\chi1}$ and $S_{\chi2}$ estimated with and
without TO have been compared.

The average correlations between $S_{\chi1}$ and $S_{\chi2}$ without
TO are $0.9816\pm0.0315$ and $0.9818\pm0.0255$ with TO. These
correlations are due to the stochastic approach in the Bayesian
framework. Correlation values are similar, indicating that the
stochastic variance has not been affected by TO.

The average cross-correlation between $S_{\chi1}$ and $S_{\chi2}$
with and without TO is $0.9760\pm0.0388$. This value is similar to
the correlation due to stochastic process, demonstrating that the TO
version is equivalent to the original version of BPSS.

No significant differences have been observed, confirming that the
TO version is equivalent to the original version of BPSS.

\section{Performance and Accuracy of CHO\label{sec:PixelSelection}}

The impact of the convex hull pixel selection pre-processing step
has been evaluated on two dataset: (i) synthetic data generated from
linear mixtures of known materials and (ii) an OMEGA hyperspectral
image of the south polar cap of Mars as an example from Planetology.
Since the BPSS with TO has been shown to be semantically equivalent
to the previous version, the TO approach is used in the rest of this
article.

\subsection{Synthetic data}

\subsubsection{Description}

Several synthetic dataset have been generated by mixing a known
number of endmembers, with abundances simulated with uniform
distribution. The generated dataset are of size $200\times500$
pixels, which is a spatial size similar to the one of a typical
hyperspectral image. For the endmembers, the following spectra have
been used: H$_{2}$O and CO$_{2}$ ice spectra
\cite{Doute_reflectancemodel_JGR1998,Schmidt_Wavanglet_IEEETGRS2007}
and mineral spectra from the USGS Digital Spectral Library splib06a
\cite{nla.cat-vn4312481}, resampled to match the $128$ wavelengths
of OMEGA C Channel \cite{Bibring:2004}. To ensure the sum-to-one
constraint on the $R$ endmember abundances, a uniform distribution
on the simplex has been used following a well established scheme
\cite{Onn}. Synthetic dataset have been generated with $3$, $5$ and
$10$ endmembers. Based on this method, dataset for which the maximum
abundance of each single endmember was limited to a certain value
($100\%$, $80\%$ and $60\%$) have also been considered. This latter
data, that are called ``cutoff" in the sequel, allows one to test
the method efficiency face to various conditions in terms of purity
of the samples (in cases where pure -- to a certain degree --
components occur in the dataset or not). In addition, a $3$
component asymmetric dataset has been investigated, with one of the
the component abundance (albite) being limited to a cutoff of $35\%$
and the abundances of the two others (ices) not being limited.
Besides, dataset with some added OMEGA-like Gaussian noise,
amplified or not, have been also generated and investigated. The
noise estimation on the dark currents of the OMEGA instruments for
observation 41\_1 has been used
\cite{Schmidt_Wavanglet_IEEETGRS2007}. Note that for all the
considered simulation scenarii, the number of sources to be
estimated has been tuned to the actual number of endmembers used to
produce the artificial dataset.

\subsubsection{Performance}

Computation times are about $50$ times shorter when pixel selection
by convex hull (CHO) is performed as a preprocessing step (see Table
\ref{Flo:TabCompTime}).

\begin{table}
\centering\begin{tabular}{|c|c|c|c|} \hline Algorithm & Without CHO
& With CHO & Time ratio\tabularnewline \hline \hline BPSS & 71463
(BPSS-2) & 2205 (BPSS-1) & 32.41\tabularnewline \hline BPSS2 &
530654 (BPSS2-2) & 7133 (BPSS2-1) & 74.39\tabularnewline \hline
\end{tabular}
\caption{Computation times (after TO) in seconds, for a synthetic
dataset with 3 endmembers (no cutoff, no noise), for both BPSS and
BPSS2, with and without CHO. In this example, $944$ pixels were
selected for the CHO, among a total of $100000$. The name of the run
of tables \ref{Flo:TableResultsArtifBPSS} and
\ref{Flo:TableResultsArtifBPSS2} is noted in parenthesis.}
\label{Flo:TabCompTime}
\end{table}

\subsubsection{Accuracy}

\paragraph{Analysis of the results}

\begin{figure}[h!]
\centering
  \includegraphics[width=\figwidth]{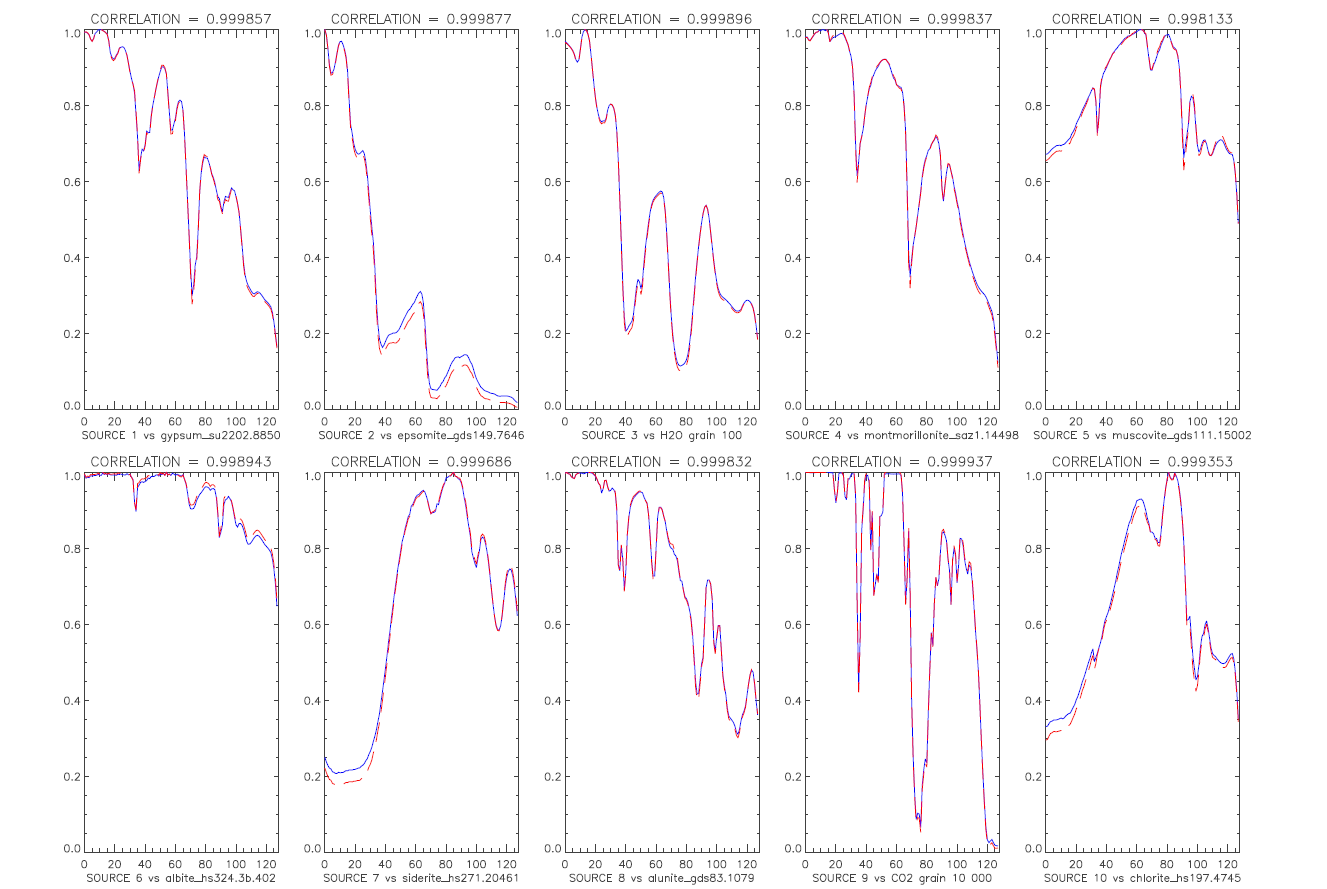}
\caption{Sources estimated by BPSS2 (blue lines) and their spectral
matches (red dotted lines), for an artificial dataset with $10$
endmembers (no cutoff, no noise).}\label{Flo:BPSS2-100-10sources}
\end{figure}

The spectrum of each estimated source has been compared to the spectra
from the spectral library containing the pure endmembers used to produce
the synthetic dataset. The absolute value of the correlation has been
used as a similarity measurement, thus as a criterion for the determination
of the best spectral match. On Figures \ref{Flo:BPSS2-100-10sources},
each source is represented along with its best match, according to
the aforementioned criterion. Table \ref{Flo:TableResultsArtifBPSS}
(resp. Table \ref{Flo:TableResultsArtifBPSS2}) shows the results
for BPSS (resp. BPSS2).

A source is considered a good estimation of a certain endmember if
both are each other best spectral match and if their absolute
correlation is greater than $80\%$. For each run, the number of
well-estimated sources is mentioned in Tables
\ref{Flo:TableResultsArtifBPSS} and
\ref{Flo:TableResultsArtifBPSS2}. Note that endmembers matched by
several sources, in case it happens, are only counted once. Along
with the number of well-estimated sources, the mean value of the
correlations between (only) the well-estimated sources and their
best spectral match also helps to the assessment of the accuracy for
the estimation of the whole set of sources for each run. Simple
distance could not be used here because the scale in usual blind
source separation is undetermined \cite{HyvarinenBOOK2001}.

\paragraph{BPSS vs. BPSS2}
In most of the tested cases, the quality of the estimation is
unambiguously better with BPSS2 than with BPSS (see Tables
\ref{Flo:TableResultsArtifBPSS} and
\ref{Flo:TableResultsArtifBPSS2}). The improvement appears to be
even more significant when the number of endmembers is increasing.
Our $3$ endmember test dataset is a mixture of two endmembers with
strong spectral signatures (CO$_{2}$ ice and H$_{2}$O ice) and a
third one with weaker signatures (albite), as often with minerals.
Interestingly, while using BPSS allows one to correctly estimate the
ices spectra but not albite, BPSS2 is actually able to correctly
estimate the three endmembers. This confirms that adding the
sum-to-one constraint is necessary when dealing with such dataset,
which is important regarding the analysis of other dataset.

\paragraph{Effect of the pixel selection (CHO)}

With the exception of the asymmetric dataset (see below), the
endmembers is less well-estimated when a pixel selection has been
performed, the loss seeming less significant when the number of
endmembers is low.

Also note that the results with pixel selection do not appear to be
very sensitive to the cutoff variations: the loss of quality
(between runs performed with and without pixel selection) is similar
for cutoffs of $60\%$, $80\%$ and $100\%$, which can be explained by
the pixel selection's ability to extract the purest available
pixels.

\paragraph{Effect of the number of endmembers}

Due to curse of dimensionality, the more endmembers to be estimated
with the fixed number of wavelength, the more difficult is the
estimation
\cite{Hughes_AccuracyDimension_IEEETrInfTheo1968,Lee_HighDim_TGRS1993}.
Still, BPSS2 gives excellent results even for $10$ sources, as all
spectra are estimated with a correlation coefficient higher than
99\% (see fig. \ref{Flo:BPSS2-100-10sources}).

\paragraph{Effect of the maximum abundance cutoff}

The cutoff affects the quality of the estimation, which is clearly
better, for BPSS and BPSS2, when pure components occur in the
dataset. This has to be remembered when dealing with real dataset.

\paragraph{Effect of noise}

\begin{figure}[h!]
\centering
  \includegraphics[width=\figwidth]{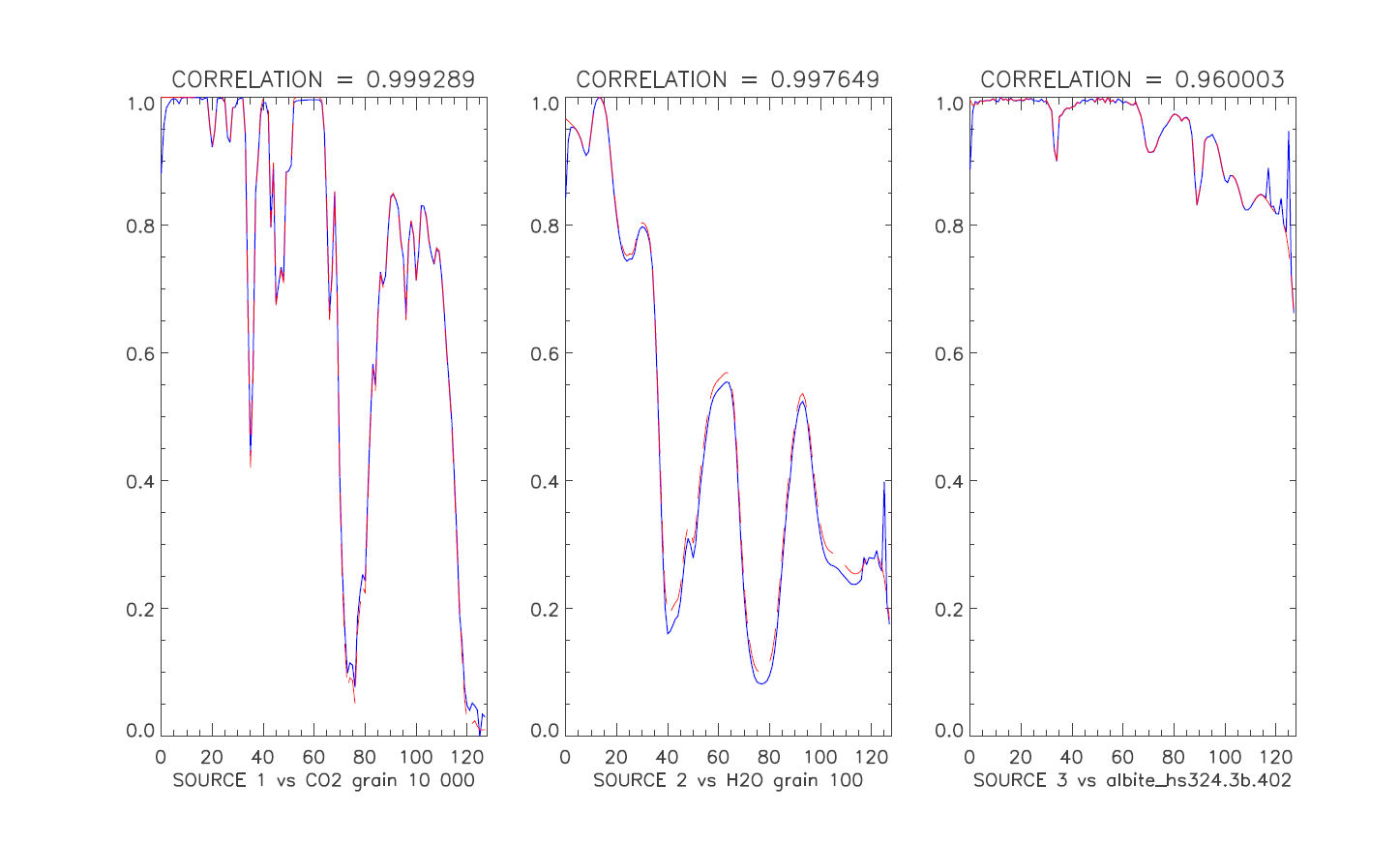}
\caption{Sources estimated by BPSS2 (blue lines) and their spectral
matches (red dotted lines), for an artificial dataset with $3$
endmembers and $100$-times amplified OMEGA-like noise (no
cutoff).}\label{Flo:BPSS2-100-3sources-100noisy}
\end{figure}

The results clearly show that the method is very robust to noise, as
the estimation of the sources does not appear to be significantly
affected by the addition of a Gaussian OMEGA-like noise to the
synthetic dataset. BPSS2 (without pixel selection) even manages to
successfully overcome the addition of a $100$-times amplified
OMEGA-like noise (see Tables \ref{Flo:TableResultsArtifBPSS2} and
\ref{Flo:BPSS2-100-3sources-100noisy}).

\paragraph{Effect of asymmetry in maximum abundance cutoff }

In this case, the results are better with pixel selection rather
than without. BPSS2 with pixel selection is the only run (performed
on this synthetic dataset) that allows one to successfully estimate
the three endmembers that have been used to generate the dataset,
including albite, whose abundances have been limited to a cutoff of
$35\%$ and whose spectral signature is weaker than the ones of the
other endmembers (ices). This result can be explained by the fact
that pixel selection is able to extract the pixels with the
strongest available albite signature, and consequently overcomes the
blinding effect of the ices occurring in the whole dataset, that has
affected the results when no pixel selection has been performed.

\begin{table*}
 \label{Flo:TableResultsArtifBPSS} \centering
\begin{tabular}{|c|c|c|c|c|c|c|}
\hline Run Id & Cutoff (\%) & Nb of endmembers & Noise & Pixel
selection & Nb of well-estimated sources & Mean correlation

(\%)\tabularnewline
\hline
\hline
BPSS-1 & 100\% & 3 & no & yes & 2/3 & 99.9441\tabularnewline
\hline
BPSS-2 & 100\% & 3 & no & no & 2/3 & 99.9963\tabularnewline
\hline
BPSS-3 & 80\% & 3 & no & yes & 2/3 & 98.5826\tabularnewline
\hline
BPSS-4 & 80\% & 3 & no & no & 2/3 & 99.9286\tabularnewline
\hline
BPSS-5 & 60\% & 3 & no & yes & 2/3 & 98.2135\tabularnewline
\hline
BPSS-6 & 60\% & 3 & no & no & 2/3 & 99.2499\tabularnewline
\hline
BPSS-7 & 100\% & 5 & no & yes & 2/5 & 92.0689\tabularnewline
\hline
BPSS-8 & 100\% & 5 & no & no & 3/5 & 92.5181\tabularnewline
\hline
BPSS-9 & 100\% & 10 & no & yes & 5/10 & 88.3081\tabularnewline
\hline
BPSS-10 & 100\% & 10 & no & no & 6/10 & 94.2996\tabularnewline
\hline
BPSS-11 & 100\% & 3 & OMEGA & yes & 2/3 & 99.9067\tabularnewline
\hline
BPSS-12 & 100\% & 3 & OMEGA & no & 2/3 & 99.9961\tabularnewline
\hline
BPSS-13 & 100\% & 3 & 10xOMEGA & yes & 2/3 & 99.8927\tabularnewline
\hline
BPSS-14 & 100\% & 3 & 10xOMEGA & no & 2/3 & 99.9976\tabularnewline
\hline
BPSS-15 & 100\% & 3 & 100xOMEGA & yes & 2/3 & 98.8182\tabularnewline
\hline
BPSS-16 & 100\% & 3 & 100xOMEGA & no & 2/3 & 98.2574\tabularnewline
\hline
BPSS-17 & ices: 100\%, alb.: 35\% & 3 & no & yes & 2/3 & 99.9811\tabularnewline
\hline
BPSS-18 & ices: 100\%, alb.: 35\% & 3 & no & no & 2/3 & 98.9855\tabularnewline
\hline
\end{tabular}
\caption{Results obtained for different synthetic dataset with the
BPSS algorithm. Characteristics of each dataset are shown: number of
endmembers, cutoff, and noise. Each dataset has been analyzed with a
number of sources to be estimated equal to the number of endmembers
used to generate the artificial dataset, with and without pixel
selection. Quality of the estimation is expressed through the number
of well-estimated sources and the mean absolute expression as
explained in the text.}
\end{table*}

\begin{table*}
\centering
\begin{tabular}{|c|c|c|c|c|c|c|}
\hline Run Id & Cutoff (\%) & Nb of endmembers & Noise & Pixel
selection & Nb of well estimated sources & Mean correlation
(\%)\tabularnewline \hline \hline BPSS2-1 & 100\% & 3 & no & yes &
3/3 & 99.8923\tabularnewline \hline BPSS2-2 & 100\% & 3 & no & no &
\textbf{3}/3 & 99.9222\tabularnewline \hline BPSS2-3 & 80\% & 3 & no
& yes & 2/3 & 95.8934\tabularnewline \hline BPSS2-4 & 80\% & 3 & no
& no & \textbf{3}/3 & 99.9200\tabularnewline \hline BPSS2-5 & 60\% &
3 & no & yes & 2/3 & 95.2965\tabularnewline \hline BPSS2-6 & 60\% &
3 & no & no & 3/3 & 97.5408\tabularnewline \hline BPSS2-7 & 100\% &
5 & no & yes & 3/5 & 99.2821\tabularnewline \hline BPSS2-8 & 100\% &
5 & no & no & \textbf{5}/5 & 99.9174\tabularnewline \hline BPSS2-9 &
100\% & 10 & no & yes & 5/10 & 98.9439\tabularnewline \hline
BPSS2-10 & 100\% & 10 & no & no & 10/10 & 99.9535\tabularnewline
\hline BPSS2-11 & 100\% & 3 & OMEGA & yes & 3/3 &
99.8726\tabularnewline \hline BPSS2-12 & 100\% & 3 & OMEGA & no &
3/3 & 99.9955\tabularnewline \hline BPSS2-13 & 100\% & 3 & 10xOMEGA
& yes & 3/3 & 99.7298\tabularnewline \hline BPSS2-14 & 100\% & 3 &
10xOMEGA & no & 3/3 & 99.9962\tabularnewline \hline BPSS2-15 & 100\%
& 3 & 100xOMEGA & yes & 2/3 & 95.4706\tabularnewline \hline BPSS2-16
& 100\% & 3 & 100xOMEGA & no & 3/3 & 98.5647\tabularnewline \hline
BPSS2-17 & ices: 100\%, alb.: 35\% & 3 & no & yes & 3/3 &
95.9402\tabularnewline \hline BPSS2-18 & ices: 100\%, alb.: 35\% & 3
& no & no & 2/3 & 99.7202\tabularnewline \hline
\end{tabular}
\caption{Results obtained for different synthetic dataset with the
BPSS2 algorithm. Characteristics of each dataset are shown: number
of endmembers, cutoff, and noise. Each dataset has been analyzed
with a number of sources to be estimated equal to the number of
endmembers used to generate the artificial dataset, with and without
pixel selection. Quality of the estimation is expressed through the
number of well-estimated sources and the mean absolute expression as
explained in the text.} \label{Flo:TableResultsArtifBPSS2}
\end{table*}

\subsection{OMEGA data}

\subsubsection{Presentation}

The OMEGA (Observatoire pour la Min\'eralogie, l'Eau, les Glaces et
l'Activit\'e) instrument is a spectrometer on board Mars Express
(European Space Agency), which provides hyperspectral images of the
Mars surface, with a spatial resolution from $300$m to $4$km, $96$
channels in the visible range and $256$ wavelength channels in the
near infra-red \cite{Bibring:OMEGA}. In this work, $184$ spectral
bands have been selected according to the best signal to noise
ratio. Conversely, spectral bands that contain the thermal emission
have been removed.

\begin{figure}[h!]
\center
\includegraphics[width=\figwidth]{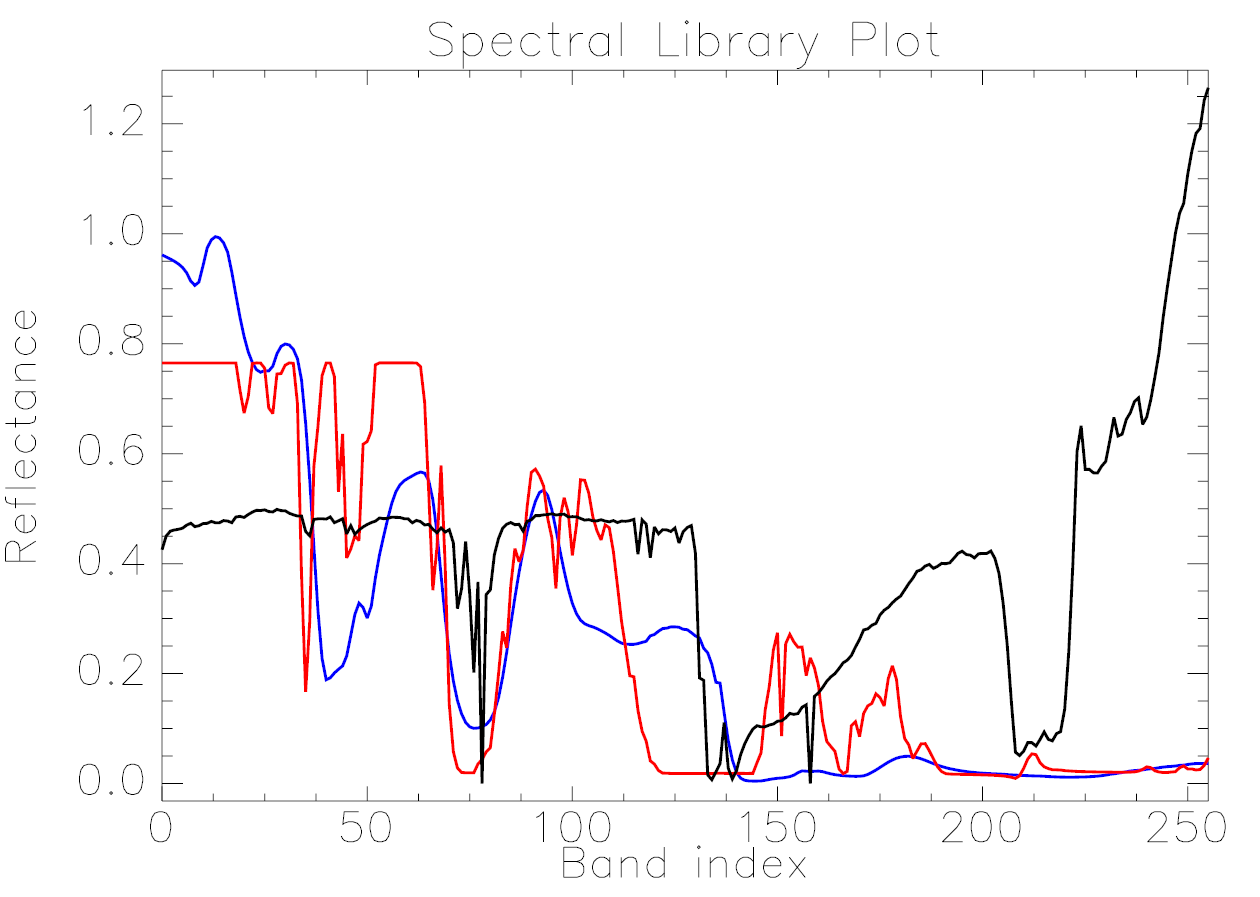}
\caption{Reference spectra of the OMEGA hyperspectral image 41\_1:
(i) in blue: synthetic H$_{2}$O ice with grain size of $100$
microns, (ii) in red: synthetic CO$_{2}$ ice with grain size of $10$
centimeters, (iii) in black: OMEGA typical dust materials with
atmosphere absorption.} \label{Flo:FigReferenceSpectra}
\end{figure}

Blind source separation on this dataset has been initiated by using
the JADE algorithm \cite{Forni2005}. In particular the image 41\_1
of the permanent south polar region has been used for supervised
classification approach with WAVANGLET
\cite{Schmidt_Wavanglet_IEEETGRS2007}, unsupervised classification
approach \cite{Galluccio_FreeGraphClustering_draft2009} and
unsupervised blind source separation using BPSS
\cite{Moussaoui_JADE-BPSS_Neurocomp2008}. Since no ground truth is
available, the results from physical non-linear inversion have been
considered as a reference
\cite{Schmidt_Wavanglet_IEEETGRS2007,Doute_PSPC_PSS2007,Bernard-Michel_MarsGRSIR_JGR2009}.
In this image, the surface is dominated by dust and some spectra
contains CO$_{2}$ and water ices (see fig.
\ref{Flo:FigReferenceSpectra}). This reference dataset for
hyperspectral classification is available online
\footnote{\url{http://sites.google.com/site/fredericschmidtplanets/Home/hyperspectral-reference}
}. The Luo \emph{et al.} method introduced in
\cite{Luo_UnsupervisedClassif_Conf2009} has estimated $2$ sources
for both 41\_1 and 41\_1.CUT images. From previous work using band
ratio detection \cite{Bibring:2004}, physical inversion of the
radiative transfer
\cite{Doute_PSPC_PSS2007,Bernard-Michel_MarsGRSIR_JGR2009},
supervised classification approach using Wavanglet
\cite{Schmidt_Wavanglet_IEEETGRS2007} and unsupervised
classification \cite{Galluccio_FreeGraphClustering_draft2009}, $3$
endmembers have been detected: dust, CO$_{2}$ and water ice. The
number of sources has been tuned to $3$ in our study.

The proportion of pixels containing CO$_{2}$ice and H$_{2}$O ice on
the 41\_1 image is estimated to be $16.76\%$ and $21.84\%$,
respectively \cite{Schmidt_Wavanglet_IEEETGRS2007}. The first $300$
lines of the 41\_1 image (subset named 41\_1.CUT) contain all
spectra containing ices. For this subset, the proportion of pixels
with detected CO$_{2}$ and H$_{2}$O is $48.72\%$ and $63.48\%$,
respecively.

\subsubsection{Performance}

Computation times are about $100$ times shorter when pixel selection
by convex hull (CHO) has been performed as a preprocessing step (see
Table \ref{Flo:TabCompTime-OMEGA}).

\begin{table}
\centering\begin{tabular}{|c|c|c|c|}
\hline
Algorithm & Without CHO & With CHO & Time ratio\tabularnewline
\hline
\hline
BPSS & 166 400 (OMEGA-5) & 1 468 (OMEGA-7) & 113.28\tabularnewline
\hline
BPSS2 & 332 680 (OMEGA-6) & 3 176 (OMEGA-8) & 104.71\tabularnewline
\hline
\end{tabular}
\caption{Computation times in seconds, for OMEGA 41\_1 image with
$3$ endmembers, for both BPSS and BPSS2, with and without CHO. In
this example, $670$ pixels have been selected with the CHO, among a
total of $111488$. The name of the run of Table \ref{tab:OMEGA} is
noted in parenthesis.} \label{Flo:TabCompTime-OMEGA}
\end{table}

\subsubsection{Accuracy}

Table \ref{tab:OMEGA} reports the results from different tests, each
run is defined by a number. To estimate the quality of the
estimation, the correlation between the reference spectra and the
estimated sources has been computed. The attribution of each source
has been done \emph{ad hoc} using both spectral source and spatial
abundances.

\paragraph{Asymmetric abundances of the sources}

The quality of estimation with both BPSS and BPSS2 is significantly
lower for dataset 41\_1 (run OMEGA-5 to OMEGA-8) in comparison with
41\_1.CUT (run OMEGA-1 to OMEGA-4). This result suggests that both
BPSS and BPSS2 are less efficient in a case of an asymmetric distribution
of the sources.

\paragraph{BPSS vs. BPSS2}

The algorithm BPSS gives significantly better results than BPSS2 (for
instance run OMEGA-3 vs OMEGA-4). This is due to non-linearity in
the radiative transfer and noise in the dataset in contradiction with
the full additivity constraint.

\paragraph{Effect of the pixel selection}

\begin{figure}[h!]
\centering
  \includegraphics[width=\figwidth]{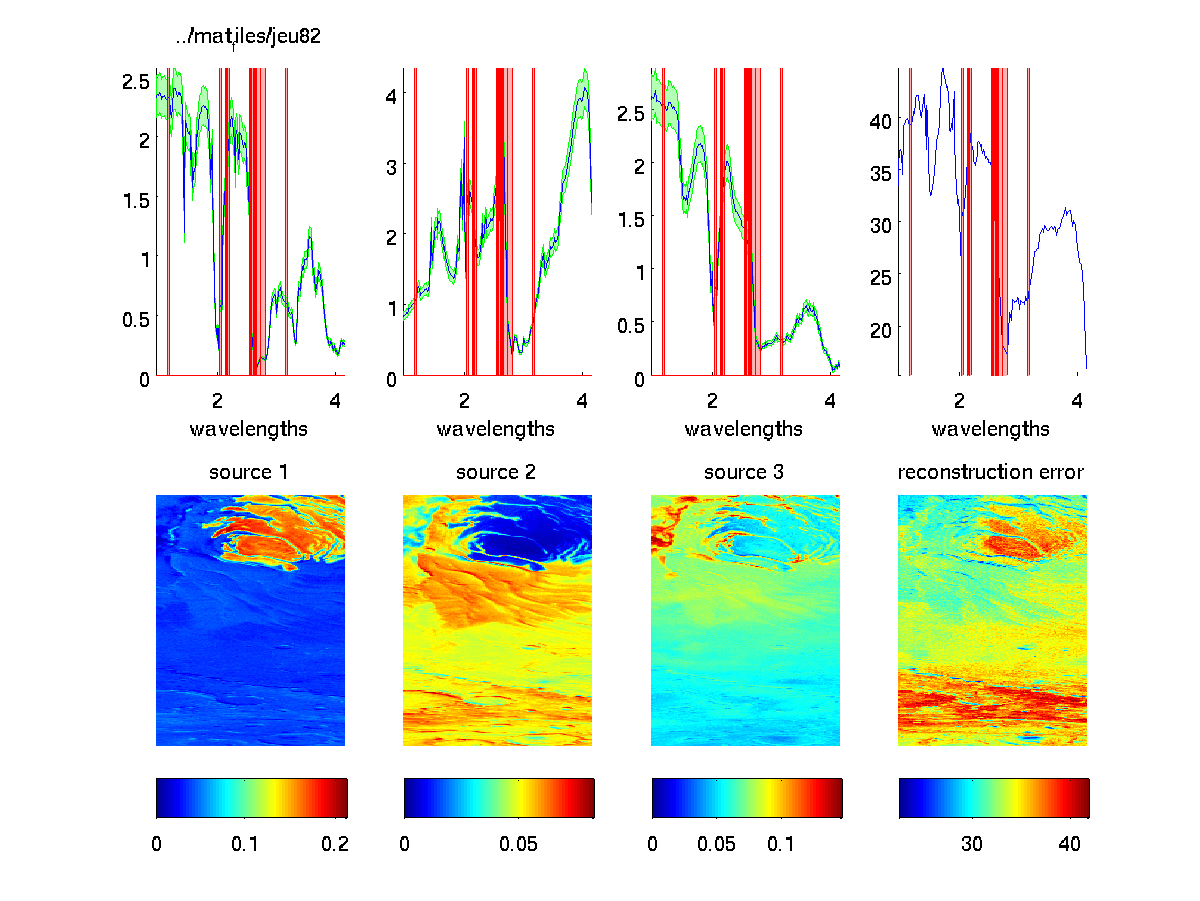}
\caption{Estimation of $3$ sources of the entire OMEGA image 41\_1
with BPSS using a preprocessing step of pixel selection using the
convex hull method. The first and third source are clearly
identified as CO$_{2}$ and H$_{2}$O ices (see fig.
\ref{Flo:FigReferenceSpectra}) with a correlation coefficient of
$0.953$ and $0.940$ (see run OMEGA-7 of Table \ref{tab:OMEGA}). The
spatial abundances is well estimated regarding the WAVANGLET
classification method
\cite{Schmidt_Wavanglet_IEEETGRS2007,Moussaoui_JADE-BPSS_Neurocomp2008}.
The second source is identified as dust with a lower correlation
coefficient ($0.372$).}\label{Flo:BPSS41_1-pixelselection}
\end{figure}

\begin{figure}[h!]
\centering
  \includegraphics[width=\figwidth]{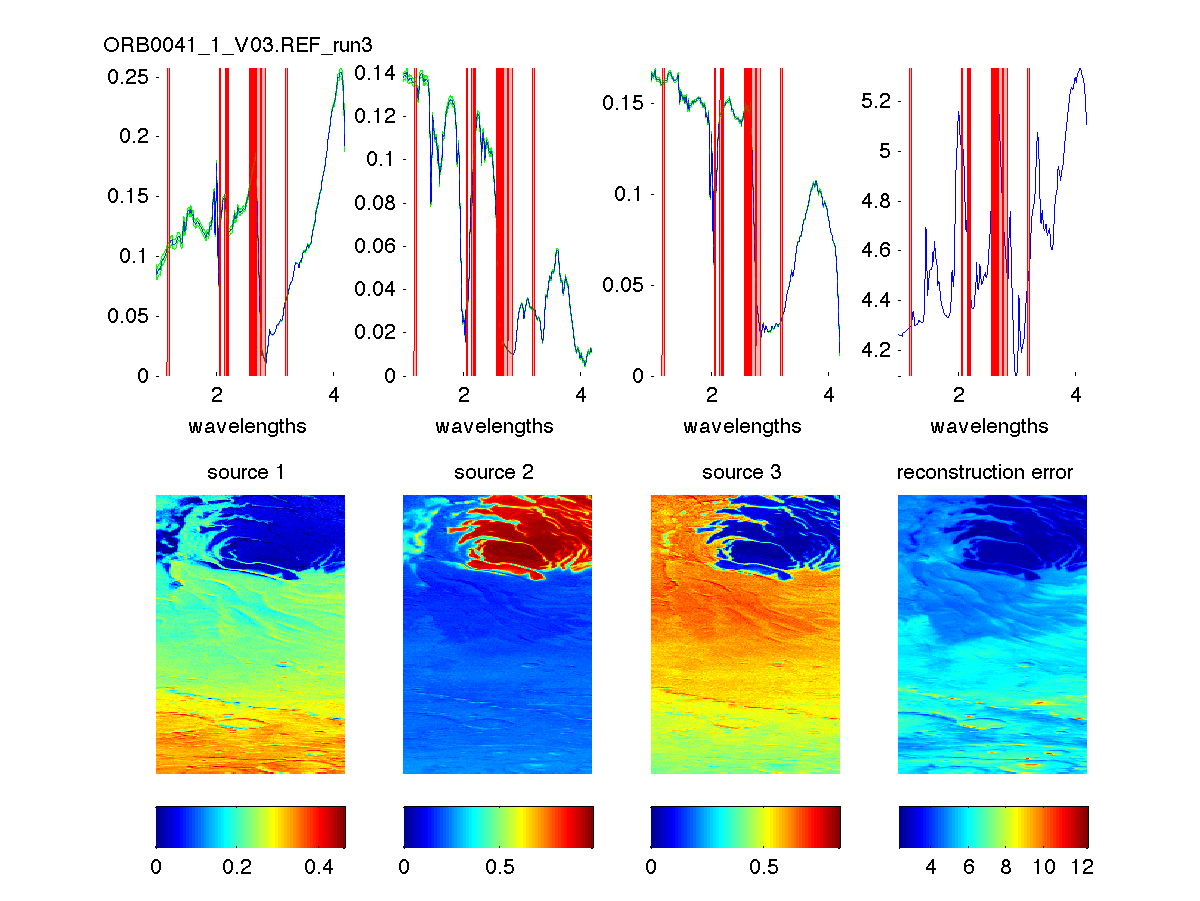}
\caption{Estimation of $3$ sources of the entire OMEGA image 41\_1
with BPSS without pixel selection. The second source is clearly
identified as CO$_{2}$ ice (see fig. \ref{Flo:FigReferenceSpectra})
with a correlation coefficient of $0.957$ (see run OMEGA-5 of Table
\ref{tab:OMEGA}). The first and third sources are identified to dust
and water ice with lower correlation coefficients of 0.555 and
0.773. The spatial abundances of water ice is not well estimated
regarding the WAVANGLET classification method
\cite{Schmidt_Wavanglet_IEEETGRS2007}.}\label{Flo:BPSS41_1-NOpixelselection}
\end{figure}

When the convex hull selection has been used as a pre-processing
step to BPSS/BPSS2, the estimation is significantly better (see fig.
\ref{Flo:BPSS41_1-pixelselection}, for run OMEGA-5 and fig.
\ref{Flo:BPSS41_1-NOpixelselection} for run OMEGA-7). These results
show that pixel selection is a way to better take into account the
occurrence of rare endmembers and thus is an interesting method to
provide better results.

\begin{table}
\centering
\begin{tabular}{|c|c|c|c|c|c|c|}
\hline \multirow{2}{*}{Run Id} & \multirow{2}{*}{Image} & \multirow{2}{*}{Algo} & pixel    & \multirow{2}{*}{H$_{2}$O}& \multirow{2}{*}{CO$_{2}$}& \multirow{2}{*}{dust}\\
                               &                        &                       &selection &  &  &   \\
\hline \hline OMEGA-1 & 41\_1.CUT & BPSS & no & 0.883 & 0.955 & 0.542\\
\hline OMEGA-2 & 41\_1.CUT & BPSS2 & no & 0.823 & 0.958 & 0.980\\
\hline OMEGA-3 & 41\_1.CUT & BPSS & yes & 0.956 & 0.951 & 0.766\\
\hline OMEGA-4 & 41\_1.CUT & BPSS2 & yes & 0.894 & 0.910 & 0.975\\
\hline OMEGA-5 & 41\_1 & BPSS & no & 0.773 & 0.957 & 0.555\\
\hline OMEGA-6 & 41\_1 & BPSS2 & no & - & 0.953 & 0.512\\
\hline OMEGA-7 & 41\_1 & BPSS & yes & 0.940 & 0.953 & 0.372\\
\hline OMEGA-8 & 41\_1 & BPSS2 & yes & 0.450 & 0.954 & 0.982\\
\hline
\end{tabular}\label{tab:OMEGA}
\caption{Results on algorithms BPSS and BPSS2 on a portion of OMEGA
image (41\_1.CUT) and on the entire image (41\_1). For 41\_1.CUT,
the proportion of pixels with detected CO$_{2}$ is $48.72\%$ and
respectively $63.48\%$ for H$_{2}$O
\cite{Schmidt_Wavanglet_IEEETGRS2007}. For 41\_1, the proportion of
pixels for CO$_{2}$ and H$_{2}$O is $16.76\%$ and $21.84\%$. The
columns H$_{2}$O, CO$_{2}$ and dust indicate the correlation
coefficient between the estimated sources and the reference spectra.
(-) indicates that no identification of H$_{2}$O neither from
spectral nor spatial results. This source has been detected to be
CO$_{2}$ ice (correlation $0.911$).}
\end{table}

\section{Discussion and conclusion\label{sec:conclusion}}

For the first time, a MCMC-based blind source separation strategy
with positivity and sum-to-one constraints has been effectively
applied on a complete hyperspectral image with a typical size
frequently encountered in Earth and Planetary Science. The
optimization of BPSS \cite{Moussaoui06tsp} and BPSS2
\cite{Dobigeon09sp} presented in this article consists of two
independent parts: (i) \emph{Technical Optimization (TO)} reduces
the memory footprint, lowers the average cost of algorithmic
operations, and makes smart re-use of memory (ii) \emph{Convex Hull
Optimization (CHO)} reduces the number of spectra to process.

\begin{figure}[h!]
\center
\includegraphics[width=\figwidth]{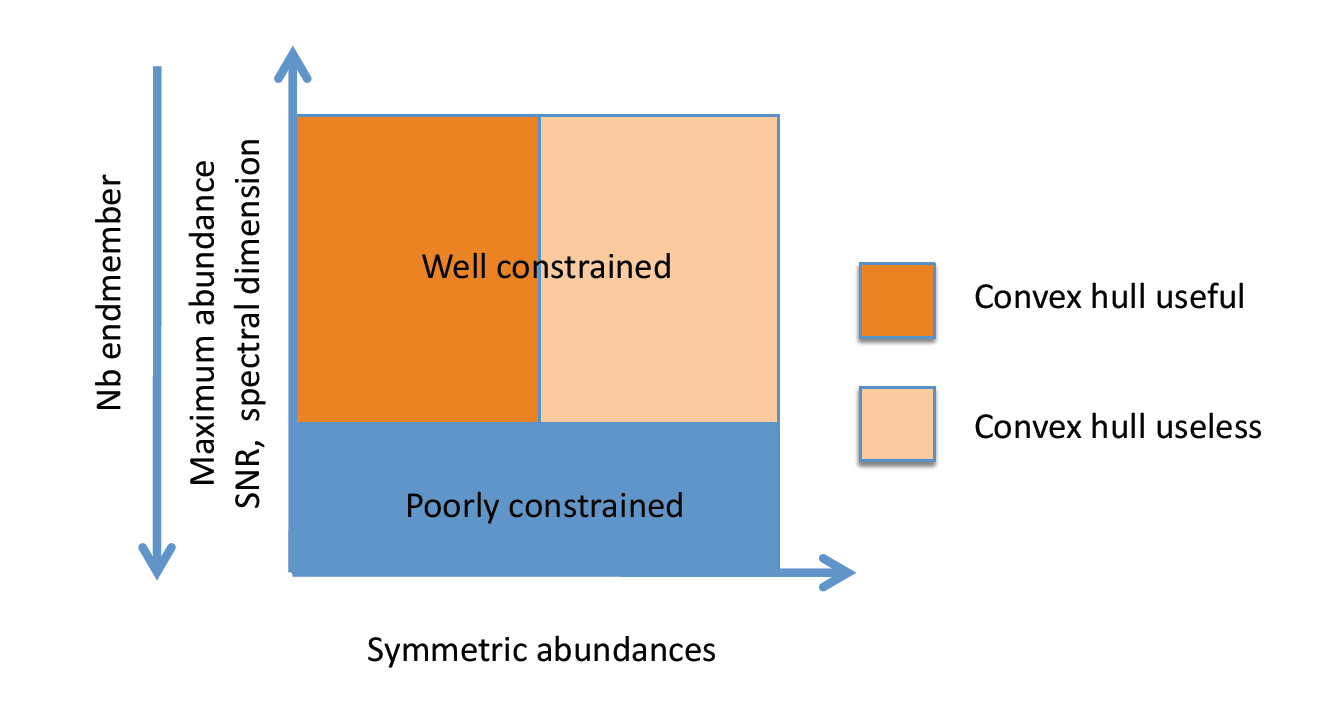}
\caption{Schematic of the source separation estimation and
usefulness of convex hull pixel selection for hyperspectral images.}
\label{Flo:SchemaSummarize}
\end{figure}

Figure \ref{Flo:SchemaSummarize} summarizes the following results
in a schematic form.
\begin{enumerate}
\item The TO, for both BPSS and BPSS2, allows one to decrease the
computation times by a factor of $1.5$, without altering the
accuracy of the results. Memory consumption has also been reduced by
a significant factor. With such unambiguous advantages, the TO
versions of BPSS and BPSS2 can be rather used than the original
implementations.
\item Trivially, results obtained for linear artificial dataset (with uniform
abundance distributions identical for each endmember with abundances
until $100\%$) have demonstrated that the sources estimated by the
TO strategy is equivalent to the CHO strategy (for instance: runs
BPSS-1 to BPSS-2 in Table \ref{Flo:TableResultsArtifBPSS} and runs
BPSS2-1 to BPSS2-2 in Table \ref{Flo:TableResultsArtifBPSS2}). In
this case, pixel selection is still relevant to reduce the
computation time about $50$ times (Table \ref{Flo:TabCompTime}).
\item Results obtained for artificial dataset with uniform abundance distributions
and identical cutoffs for all endmembers have shown that the
estimation of the sources is less accurate when a pixel selection
(CHO) has been performed (runs BPSS-3 to BPSS-6 in Table
\ref{Flo:TableResultsArtifBPSS} and runs BPSS2-3 to BPSS2-6 in Table
\ref{Flo:TableResultsArtifBPSS2}). In this case, despite $50$ times
shorter computation times, using pixel selection as a preprocessing
step seems to be inadequate.
\item For OMEGA data, the computation time reduction due to CHO has been around
$100$ (Table \ref{Flo:TabCompTime-OMEGA}). Abundance distributions
can be significantly unbalanced (some endmembers are significantly
less present in the scene). In that case, pixel selection by convex
hull (CHO) is a way to overcome the bias caused by the overwhelming
endmembers. This has been supported by the results obtained for the
synthetic dataset of linear mixture using unbalanced uniform
distribution.
\item BPSS2 seems to better estimate the sources in the artificial dataset
but not in the real case. This is probably due to non-linearity or
non-Gaussian noise effect.
\item The method BPSS2 appears to be very robust to Gaussian noise, as shown
by the results obtained on synthetic dataset, even with $100$ times
actual OMEGA noise.
\item Sometimes, some sources have been well estimated but anti-correlated with
the real spectra. This behavior has been interpreted to be due to
linear dependent endmembers. In that case, spectra built by a linear
combination of all sources except the considered source already
contain spectral signatures of the considered source. The last
source is then anti-correlated with the corresponding endmember to
decrease his contribution. This behavior has to be studied in
further details because it is clearly a limitation of blind source
separation.
\end{enumerate}
In the future, the choice of the number of sources, which is an
input in the current implementation, should be automated to allow
one batch processing without human intervention. A methodology of
pixel selection for use across dataset should also be established to
enables integration of source separation techniques into larger
systems and aim at the generation of catalogs and maps.

\section{Acknowledgments}

The authors would like to thank J. P. Bibring and the OMEGA Team for
providing the OMEGA dataset. They are also grateful to S. Dout\'e
and B. Schmitt for their ice spectral library. They acknowledge
support from the Faculty of the European Space Astronomy Centre
(ESAC). The authors finally thank two anonymous reviewers for their
valuable comments and suggestions that enabled significant
improvements to this paper.

\bibliographystyle{IEEEtran}
\bibliography{Schmidt_IEEE_Trans_GRS_2010}

\begin{IEEEbiography}[{\includegraphics[width=1in,height=1.25in,clip,keepaspectratio]{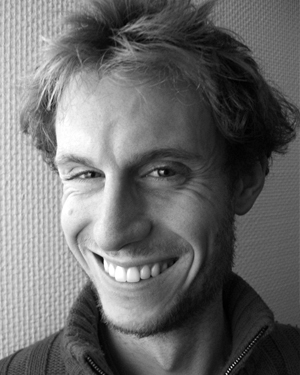}}]{Fr\'ed\'eric Schmidt}
received his Ph.D. in 2007 at the Laboratoire de Plan\'etologie de
Grenoble (CNRS-UJF), Grenoble, France. He spent two years at
European Space Agency (ESAC, Madrid) as a post-doctoral fellow.
Since 2009, he has been assistant professor at the laboratory
Interaction et Dynamique des Environnements de Surface (Universit\'e
Paris Sud - CNRS). His research interests are analysis of
hyperspectral data, ices and polar processes on planet Mars. He is
Co-Investigator of the OMEGA imaging spectrometer onboard Mars
Express (ESA).
\end{IEEEbiography}

\begin{IEEEbiography}[{\includegraphics[width=1in,height=1.25in,clip,keepaspectratio]{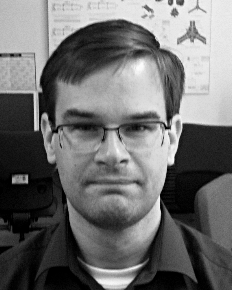}}]{Albrecht Schmidt}
received his Ph.D. in 2002 from the University of Amsterdam.  After
working three years at the University of Aalborg (Denmark), he
joined the European Space Agency, where he now works as a Computer
Scientist in Solar System Operations Division.  His research
interests are data management and data analysis.
\end{IEEEbiography}
\vfil \vfill

\begin{IEEEbiography}[{\includegraphics[width=1in,height=1.25in,clip,keepaspectratio]{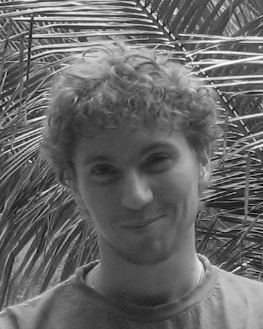}}]{Erwan Tr\'eguier}
received the Engineer Degree from the \'Ecole Nationale Sup\'erieure
d'Ing\'enieurs en Constructions A\'eronautiques (ENSICA, now part of
the Institut Sup\'erieur de l'A\'eronautique et de l'Espace,
Toulouse, France) in 2001. He received the Ph.D Degree from the
Universit\'e Paul Sabatier (UPS, Toulouse, France) in 2008, after
defending his thesis at the Centre d'\'Etude Spatiale des
Rayonnements (CESR). He is currently holding a postdoctoral position
at ESAC. His main field of research is planetary science. His
research interests include the composition of the Martian surface,
from both in-situ and orbital data, geochemical modeling of
alteration, and investigating multidimensional dataset through
statistical approaches.
\end{IEEEbiography}

\begin{IEEEbiography}[{\includegraphics[width=1in,height=1.25in,clip,keepaspectratio]{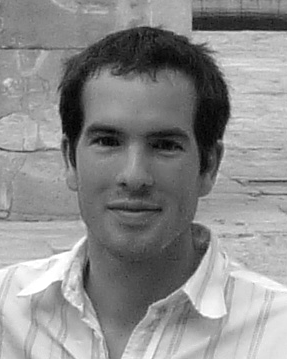}}]{Ma\"el Guiheneuf}
received the M.S. degree in Signal Processing in 2009 from \'Ecole
Centrale de Nantes (France). He was a student fellow at ESAC during
six months in 2009 on the following subject: ``Optimization
strategies for hyperspectral unmixing using Bayesian source
separation".
\end{IEEEbiography}

\begin{IEEEbiography}[{\includegraphics[width=1in,height=1.25in,clip,keepaspectratio]{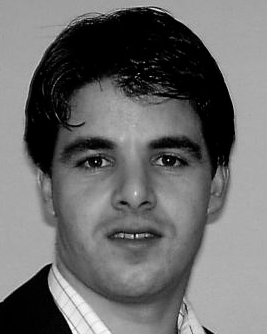}}]{Sa\"id Moussoui}
received the State engineering degree from \'Ecole Nationale
Polytechnique, Algiers, Algeria, in 2001, and, in 2005, the Ph.D.
degree in Automatic Control and Signal Processing from Universit\'e
Henri Poincar\'e, Nancy, France.

He is currently an Associate Professor at \'Ecole Centrale de
Nantes. Since September 2006, he is with the Institut de Recherche
en Communication et Cybern\'etique de Nantes (IRCCYN, UMR CNRS
6597). His research interests are in statistical signal and image
processing including source separation, Bayesian estimation and
their applications.
\end{IEEEbiography}

\begin{IEEEbiography}[{\includegraphics[width=1in,height=1.25in,clip,keepaspectratio]{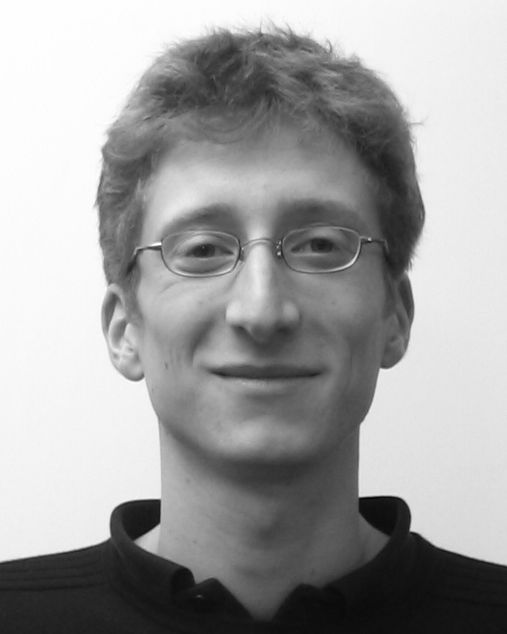}}]{Nicolas Dobigeon}
(S'05--M'08) was born in Angoul\^eme, France, in 1981. He received
the Eng. degree in electrical engineering from ENSEEIHT, Toulouse,
France, and the M.Sc. degree in signal processing from the National
Polytechnic Institute of Toulouse, both in 2004. In 2007, he
received the Ph.D. degree in signal processing also from the
National Polytechnic Institute of Toulouse.

From 2007 to 2008, he was a postdoctoral research associate at the
Department of Electrical Engineering and Computer Science,
University of Michigan. Since 2008, he has been an Assistant
Professor with the National Polytechnic Institute of Toulouse
(ENSEEIHT - University of Toulouse), within the Signal and
Communication Group of the IRIT Laboratory. His research interests
are centered around statistical signal and image processing with a
particular interest to Bayesian inference and Markov chain Monte
Carlo (MCMC) methods.
\end{IEEEbiography}

\vfil \vfill

\end{document}